\documentclass[aps,pre,twocolumn,superscriptaddress,showpacs,showkeys]{revtex4-1}
\usepackage{amsmath,amssymb, amsthm}
\usepackage{graphicx,epstopdf}
\usepackage{color}
\usepackage{subfigure}
\usepackage{eufrak}
\usepackage{hyperref}
\usepackage{enumitem}
\usepackage{xcolor}
\usepackage{xurl}
\usepackage{mciteplus}
\usepackage[english]{babel}
\usepackage[normalem]{ulem}
\usepackage{booktabs}
\usepackage{float}
\usepackage{comment}
\usepackage{chngcntr}
\usepackage{rotating}
\usepackage{soul}
\usepackage{multirow}
\renewcommand\harvardurl[1]{\textbf{URL}: \url{#1}} 
\usepackage{xurl} 

\theoremstyle{definition}
\newtheorem{definition}{Definition}[section]

\definecolor{redcolor}{rgb}{0,0,0}

\definecolor{pinkcolor}{rgb}{0,0,0}
\def\Pink#1{{\color{pinkcolor} #1}}

\definecolor{bluecolor}{rgb}{0,0,0}
\def\Blue#1{{\color{bluecolor} #1}}

\definecolor{greencolor}{rgb}{0,0,0}
\def\Green#1{{\color{greencolor} #1}}

\begin{document}
\title{Comparative analysis of stationarity for Bitcoin and the S\&P500 }

\author{Yaoyue Tang}
    \affiliation{Modelling and Simulation Research Group, The University of Sydney, Sydney NSW 2006, Australia}
    
\author{Karina Arias-Calluari}
    \affiliation{School of Mathematics and Statistics, The University of Sydney, Sydney, NSW 2006, Australia}
    
\author{Michael S. Harr\'e}
    \affiliation{Modelling and Simulation Research Group, The University of Sydney, Sydney NSW 2006, Australia}

\begin{abstract}
    This paper compares and contrasts stationarity between the conventional stock market and cryptocurrency. The dataset used for the analysis \textcolor{black}{is the} intraday price indices of the S\&P500 from 1996 to 2023 and \textcolor{black}{the} intraday Bitcoin indices from 2019 to 2023, \Green{both in USD}. We adopt the definition of `wide sense stationary', which \textcolor{black}{constrains} the time independence of the first and second moments of a time series. The testing method used in this paper follows the Wiener-Khinchin Theorem, \textcolor{black}{i.e., that for a wide sense stationary process, the power spectral density and the autocorrelation} are a Fourier transform pair. We demonstrate that localized stationarity can be achieved by truncating the time series into segments, and for each segment, detrending and normalizing the price return are required. \textcolor{black}{These results show} that the S\&P500 price return can achieve stationarity for the full 28-year period with a detrending window of 12 months and a constrained normalization window of 10 minutes. With truncated segments, a larger normalization window \textcolor{black}{can be used to establish stationarity, indicating that within the segment the data is more homogeneous}. For Bitcoin price return, the segment with higher volatility presents stationarity \textcolor{black}{with} a normalization window of 60 minutes, whereas stationarity cannot be established in other segments.
\end{abstract}


\maketitle
\section{Introduction}
In this study, \textcolor{black}{we apply a novel method to the comparative analysis of the stationarity of two univariate time series: The intraday S\&P500 and the intraday Bitcoin index in USD (BTC/USD). Stationarity plays a central role in much of stochastic time series analysis, see for example prediction and non-linearity~\cite{cheng2015time,aue2015prediction} and the distinction between varieties of stationarity~\cite{jentsch2015test,dahlhaus2012locally}. In the absence of stationarity, temporally localized dynamics can be informative of important market dynamics, such as volatility~\cite{bhowmik2020stock,harvey2016tests,jin2017time} and other forms of market instability~\cite{onnela2003asset,onnela2003dynamic,onnela2003dynamics,harre2009phase,harre2015entropy,harre2022entropy,harre2023detecting}. In this sense the study and estimation of the stationarity of market dynamics have played an important role in broadening our understanding of these dynamics. Of particular interest are the transformations that translate non-stationary processes into some form of stationarity~\cite{gardner1978stationarizable}, e.g. strict-sense, wide-sense, or locally stationary process.}

\Blue{Univariate time series found in real life are often heterogeneous and experience different types of nonstationarities \cite{Chen2002effect}. The segmenting and detrending process have been widely used to eliminate the effects of non-stationarity. However, verifying whether the time series is stationary after detrending requires more robust tests. One widely applied test for stationarity is the Dickey-Fuller test, which is suitable for autoregressive time series with normal random noise. Financial time series have been reported to be represented by an autoregressive model but with a q-Gaussian noise \cite{Tsallis1988Possible, borland2002option}}\Green{, which makes the Dickey-Fuller test less ideal. This paper explores alternative approaches for testing stationarity based on the Wiener-Khinchin Theorem, which states the that the power spectrum of a stationary random process is the Fourier transform of its autocorrelation function \cite{witt1998testing}. While the problem of segmenting a time series to remove the trend has been addressed before, similar analyses have led to different results due to variations in their optimisation processes \cite{last2008detecting}. By examining different optimal time windows, we aim to provide a more comprehensive understanding of their effectiveness and applicability, especially in financial time series}. \par
{The Wiener-Khinchin Theorem method has been applied in different settings such as geophysical \cite{witt1998testing, vincent2010resolving} and physiological data \cite{witt1998testing, hegger2000coping}. More recently, it has been applied in finance and economics to analyze the exchange rates \cite{lim2023long} and the stock market time series \cite{arias2022testing}. In this paper, we apply this novel method to both the traditional stock market (the S\&P500) and Bitcoin cryptocurrency, filling the research gap regarding the application of this method in non-Gaussian distributed time series.} \textcolor{black}{In order to introduce the approach (please see our previous work~\cite{arias2018closed,alonso2019q,arias2021methods,arias2022testing,tang2024stylized} for a more detailed introduction to the markets and recent results on non-linear analysis) we first introduce the notion of stationarity we will be working with and then discuss the approach we have taken. We first recall the following definitions that will be useful to us:}
\textcolor{black}{\begin{definition}[Strictly Stationary Process]
A continuous time random process $\{X_t\}$ is a {\it Strictly Stationary Process} (SSP) if, for a finite integer $n$, subscripts $t_1, t_2, \ldots, t_n$, and a cumulative distribution function $F_X(x_{t_1}, x_{t_2}, \ldots, x_{t_n})$, the following property holds: $$F_X(x_{t_1+\tau}, \ldots, x_{t_n+\tau}) = F_X(x_{t_1}, \ldots, x_{t_n})$$ 
for all $\tau, t_1, \ldots, t_n \in \mathbb{R} \text{ and for all } n \in \mathbb{N}_0  \quad \blacksquare$
\end{definition}}

\textcolor{black}{\noindent This definition is often too restrictive, accounting as it does for all moments of the distribution. Instead a more tractable definition is often used, accounting only for a subset of the moments, one of the most commonly retaining equality only of the first two moments:
\begin{definition}[Wide-Sense Stationary Process]
A continuous time random process $\{X_t\}$ (as above) is a {\it Wide-Sense Stationary Process} (WSP) if it constrains the mean $\mu_X(t) \triangleq \mathbb{E}[X_t]$ and  auto-covariance $K_{XX}(t_1,t_2) \triangleq \mathbb{E}[(X_{t_1} - \mu_X(t_1))(X_{t_2} - \mu_X(t_2))]$ in the following fashion:
\begin{align*} 
\mu_X(t) &= \mu_X(t + \tau) & \text{for all } \tau, t \in \mathbb{R} \\ 
K_{XX}(t_1, t_2) &= K_{XX}(t_1 - t_2, 0) & \text{for all } t_1, t_2 \in \mathbb{R} \\
\mathbb{E}[\|X_t\|^2] &< \infty & \text{for all } t \in \mathbb{R} \quad \blacksquare
\end{align*}
\end{definition}}

\textcolor{black}{In this work, we use the WSP definition to establish the transformations needed to be applied to $\{X_t\}$ (defined in the following section) to establish stationarity of the time series via the Wiener-Khinchin Theorem. The purpose of this analysis is to compare and contrast the local stationarity of the S\&P500 and BTC/USD. This is central to improving our understanding of the extent to which the dynamics reflect qualities desirable of a mature asset market that is not dominated by speculation or bubbles and crashes.}

\textcolor{black}{Our approach broadly follows that of St\u{a}ric\u{a} and Granger~\cite{stuaricua2005nonstationarities}, in their study they considered the inter-day S\&P500 statistics by ``identifying the intervals of homogeneity, that is, the intervals where a certain estimated stationary model describes the data well.'' In that approach, parameters are time-dependent co-factors to be estimated within each temporal interval. In its specifics, though, our approach differs somewhat from that of St\u{a}ric\u{a} and Granger.} \Blue{Our alternative looks at the time series at a higher resolution.} \textcolor{black}{We use intra-day price return $\{X_t\}$ and consider what (fixed) size of a moving window over which the mean and standard deviation are estimated is necessary to bring the entire time series into a stationary state, this contrasts with local parameter estimation for establishing locally stationary intervals in St\u{a}ric\u{a} and Granger.} 

\textcolor{black}{This novel extension of previous methods localizes the stationarity constraint in the sense that the entire series is then stationary, but this stationarity depends on a specific time windowing of parameter estimation rather than using the entire time series or a large time window defining an estimation interval of years. In effect, the narrower this window, the more temporally unstable the time series, and in principle, there is no obvious requirement that any fixed estimation window of, for example, the variance should necessarily be adequate to bring the entire time series into a stationary state. Ultimately, we find that we still do need to consider multiple (but quite large) partitions of the whole times series, but we also find that in some cases, it works for the entire time series, and in others, it simply is not possible to establish stationarity.}

\section{Methodology}
We recall the definitions for the governing equations of the stock market \cite{arias2022testing}, starting with defining the price return at time $t$:
\begin{equation}
X(t_{0},t)=I(t_{0}+t)-I(t_{0}),
\label{eq:Price_return}
\end{equation}
where $I(t_{0})$ is the stock market index at time $t_{0}$, and $I(t)$ is the stock market index for any time  $t > t_{0}$.

It is assumed that the stock market index $I(t)$ can be decomposed into two time series, consisting of a deterministic trend $\overline{I}(t)$ and a stochastic noise $I^*(t)$: 
\begin{equation}
    I(t) = \overline{I}(t) + I^*(t).
\end{equation}
Then, the price return $X(t)$ can also be decomposed into a deterministic component $\overline{X}(t)$ and a stochastic noise $x(t)$, which has been \textcolor{black}{modelled as} $q$-Gaussian noise (\textcolor{black}{see~\cite{arias2018closed,alonso2019q}}):
\begin{equation}
X(t) = \overline{X}(t) +x(t),
\label{eq:detrended}
\end{equation}
where $\overline{X}(t)= \overline{I}(t_0+t)-\overline{I}(t_0)$ and $x(t)= I^*(t_0+t)- I^*(t)$. 

We perform the analysis on the price returns for both markets, which includes a detrending process to eliminate the mean, and a regularization process to eliminate the standard deviation. We use the Moving Average for detrending, where a sliding window of size $\Delta_1 t$ is shifted forward along the time series (with total length $N$), and the arithmetic average of the index $I(t)$ from $[t, \,\,\, t+\Delta_1 t]$ is calculated as the trend at time $t$. For the windows at the beginning and the end of the time series where $t<\frac{\Delta_1 t}{2}$ and $t>N-\frac{\Delta_1 t}{2}$, we identify three segments for calculating the moving window with respect to the window position:

\begin{description}[font=$\bullet$~\normalfont]
\item [For $t< \frac{\Delta_1 t}{2}$] 
\begin{equation}\label{eq:MA1}
\overline{I}(t)=\frac{2}{\Delta_1 t+2t}{\sum_{k=-t+1}^{\lceil(\Delta_1 t-1)/{2}\rceil} I(t+k)},
\end{equation}
\item [For $\frac{\Delta_1 t}{2}<t<N-\frac{\Delta_1 t}{2}$] 
\begin{equation}\label{eq:Ma2}
\overline{I}(t)=\frac{1}{\Delta_1 t}{\sum_{k=-\lfloor(\Delta_1 t-1)/{2}\rfloor}^{\lceil(\Delta_1 t-1)/{2}\rceil} I(t+k)},
\end{equation}
\item [For $t>N-\frac{\Delta_1 t}{2}$] 
\begin{equation}\label{eq:MA3}
\overline{I}(t)=\frac{2}{2N-2t+\Delta_1 t}{\sum_{k=-\lfloor(\Delta_1 t-1)/{2}\rfloor}^{\lceil N-t\rceil} I(t+k)},
\end{equation}
\end{description}
with the time step of $t=1,2,3....N$ for the index fluctuations.

From the detrended price return, \Green{$x^*(t)$}, a normalization process is performed to obtain the normalized price return, \Green{$x^{**}(t)$}, given by \textcolor{black}{the standard score}:
\begin{equation} \label{regularised}
    x^{**}(t) =\frac{x^* (t)}{\sigma (x^*)},
\end{equation}
where $\sigma (x^*)$ is the standard deviation of the time series. A sliding window is shifted forward for calculating $\sigma (x^*)$, following the same segmenting technique introduced in eq. \ref{eq:MA1}-\ref{eq:MA3}. The size of this sliding window is denoted as $\Delta_2 t$ for this calculation.

Based on previous results \cite{arias2022testing}, the stationarity test is conducted on the standard score of the price return, $x^{**}(t)$. We first calculate sample autocorrelation of a time series as \cite{box2015time, liu1999statistical}:
\begin{equation}
    C_{x}(s)=\frac{\langle x(t)\, x(t+s) \rangle}{ \sigma^2_x},
\end{equation}
where $x(t)$ denotes the time series used for calculation, $s$ is the time lag, and $\sigma^2_x=\langle (x(t)-\mu)^2 \rangle$. In this analysis, the autocorrelation is calculated on the standard score of the price return by replacing $x(t)$ by \Green{$x^{**}(t)$}. 

The Fourier transform of a time series $x(t)$ is calculated as:
\begin{equation}
    \widehat{x}(f)=\dfrac{1}{\sqrt{T}}\sum_{t=0}^{T} x(t) e^{-2 \pi i f t} .
\end{equation}
The method for the stationarity test follows the Wiener–Khinchin theorem, where for a wide-sense-stationary stochastic process, the Fourier transform of the autocorrelation ($\widehat{C_x}(f)$) of a time series equals its power spectral density (PSD), denoted as $|\widehat{x}(f)|^2$:
\begin{equation}
    |\widehat{x}(f)|^2 = \widehat{C_x}(f).
\end{equation}

A smoothing technique of the \textcolor{black}{Savitzky-Golay filtering~\cite{schafer2011savitzky}} is applied on both $\widehat{C_x}(f)$ and $|\widehat{x}(f)|^2$ in order to reduce the noise in the data without affecting their trends. The smoothed result is based on the least squares of a linear fitting across a moving window, applying the normalized convolution integers of Savitzky-Golay, denoted as $C_i$. Here, the smoothing window size is chosen as $3.97 \times 10^{-5}$ Hz following a previous study \cite{arias2022testing}.

\section{Data and Results}
The dataset used in this study includes the S\&P500 price index and BTC/USD exchange rate, both reported at one-minute intervals. The S\&P500 index data cover the period from Jan 1st, 1996 to Dec 31st, 2023, and Bitcoin data span from Apr 2nd, 2019 to Dec 31st, 2023. We conduct a stationarity test on the S\&P500 time series first before moving to Bitcoin data. Both detrending and normalizing processes are required for each dataset before the stationarity test can be performed. 

\begin{figure}[!ht]
\centering
\includegraphics[scale=0.5, clip]{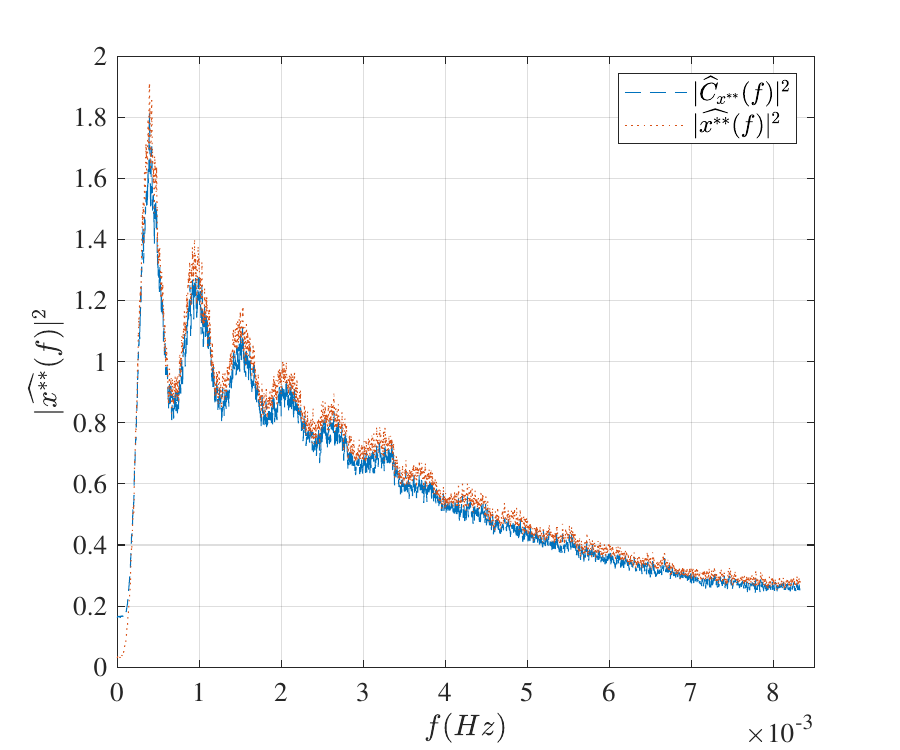}
\caption{Reproduced results with the S\&P500 price return Jan 1996 to March 2021, detrended with a 1-year window, and normalized with a 1-hour window. The PSD aligns with the Fourier transform of autocorrelation, presenting stationarity.}
\label{fig:Kari_result_rep}     
\end{figure}

The initial step is to replicate the stationarity test result for the S\&P500 normalized price return in the previous study, covering the sample period from Jan 1996 to March 2021. A detrending time window of $\Delta_1 t=1$ year is used, then the detrended price return is normalized with a regularization window $\Delta_2 t= 60$ mins to arrive at the normalized price return, \Green{$x^{**}(t)$}. Figure \ref{fig:Kari_result_rep} presents the result of the stationarity test, with the blue curve representing the Fourier transform of autocorrelation, and the orange curve for PSD. A match between $|\widehat{x^{**}} (f)|^2$ and $\widehat{C_{x^{**}}} (f)$ is achieved for time lags larger than 30 min, indicating stationarity.

\begin{figure}[!ht]
\centering
\includegraphics[scale=0.6, clip]{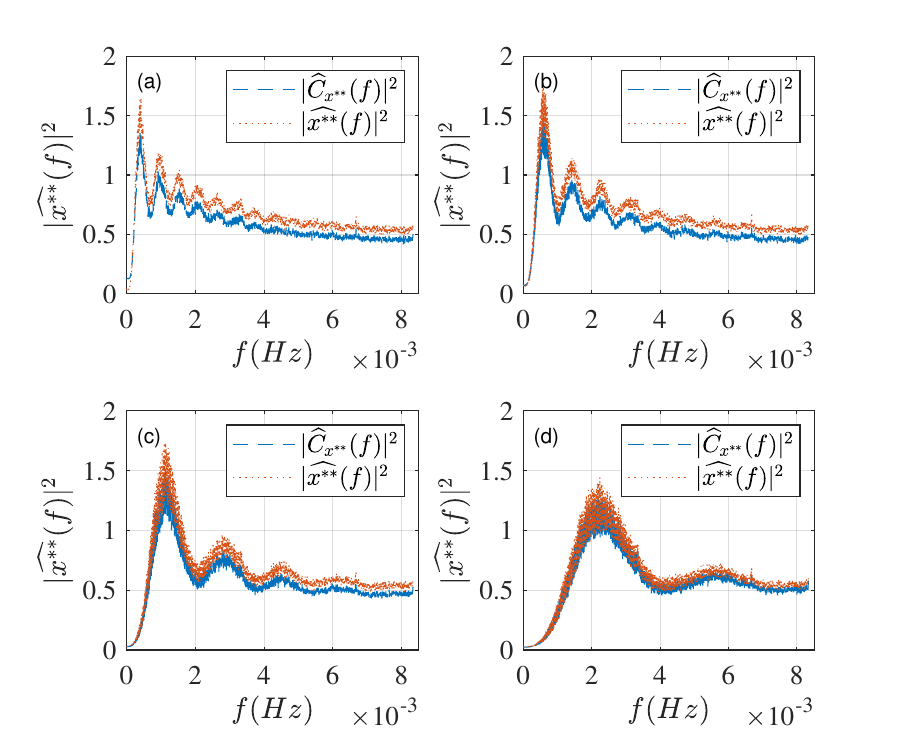}
\caption{Stationarity test with S\&P500 data from Jan 1st, 1996 to Dec 31st, 2023. (a) regularization window of 60 min. (b) regularization window of 40 min. (c) regularization window of 20 min. (d) regularization window of 10 min.}
\label{fig:SPX1996_2023}     
\end{figure}

We expanded the dataset to include the more recent time period from April 2021 onwards. Applying the same parameters with $\Delta_1 t=1$ year, and $\Delta_2 t= 60$ mins, the testing results are illustrated in Figure \ref{fig:SPX1996_2023}-a. It can be observed that $|\widehat{x^{**}} (f)|^2$ and $\widehat{C_{x^{**}}} (f)$ appear as parallel curves, indicating that stationarity does not hold with these parameters. Attempts were made to narrow the detrending time window $\Delta_1 t$ to 1 week while keeping $\Delta_2 t= 60$ mins; however, stationarity still could not be achieved. Figures \ref{fig:SPX1996_2023}-b, \ref{fig:SPX1996_2023}-c, and \ref{fig:SPX1996_2023}-d presents the test results with a gradual decline in the regularization window $\Delta_2 t$, which takes values of 40 min, 20 min, and 10 min, respectively. The gap between $|\widehat{x^{**}} (f)|^2$ and $\widehat{C_{x^{**}}} (f)$ decreases with a gradual decrease in the size of $\Delta_2 t$, indicating an increase in the degree of stationarity. The normalized price return achieves stationarity with $\Delta_2 t=10$ min for the full data set of the S\&P500 from Jan 1996 to Dec 2023. 


\begin{figure}[!ht]
\centering
\includegraphics[scale=0.6, clip]{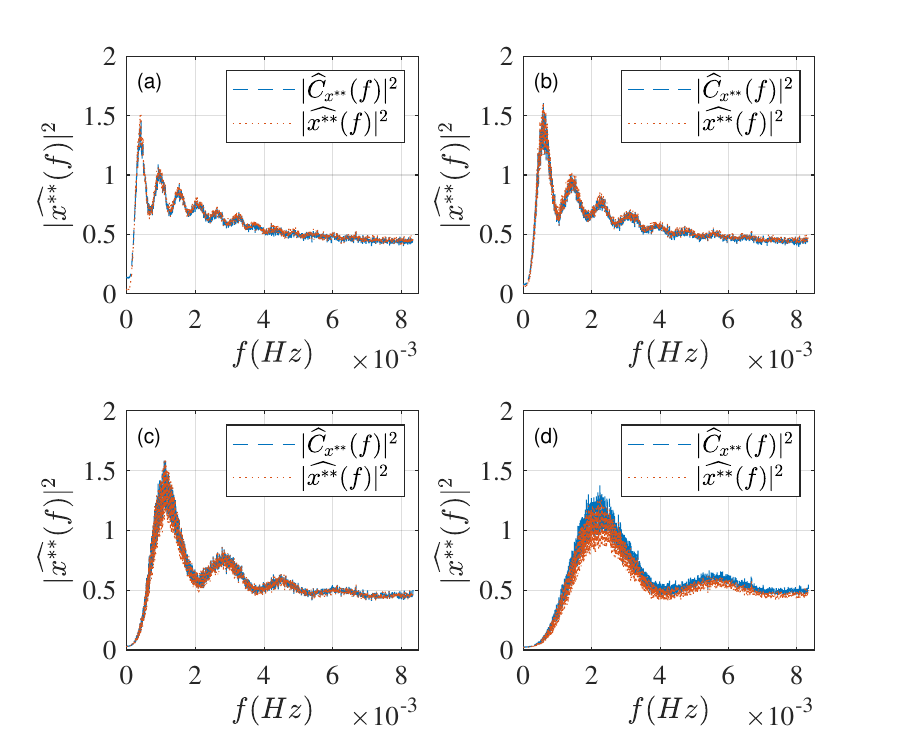}
\caption{Stationarity test with S\&P500 data from Jan 1st, 1996 to Feb 21st, 2020. (a) regularization window of 60 min. (b) regularization window of 40 min. (c) regularization window of 20 min. (d) regularization window of 10 min.}
\label{fig:SPX1996_2020}     
\end{figure}

\begin{figure}[!ht]
\centering
\includegraphics[scale=0.6, clip]{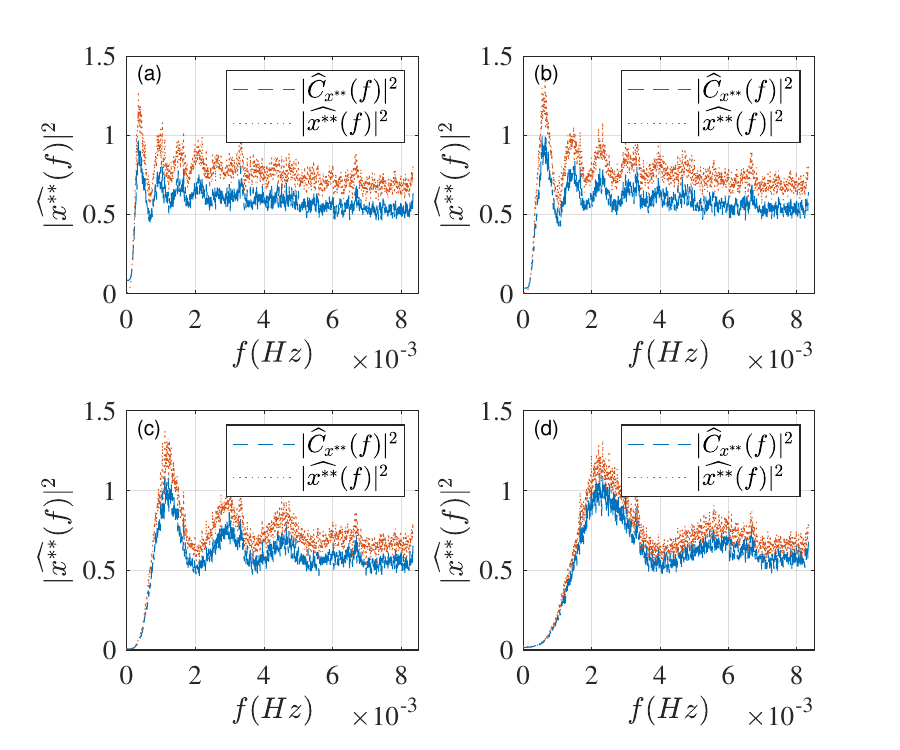}
\caption{Stationarity test with S\&P500 data from Feb 24th, 2020 to Dec 31st, 2023. (a) regularization window of 60 min. (b) regularization window of 40 min. (c) regularization window of 20 min. (d) regularization window of 10 min.}
\label{fig:SPX2020_2023}     
\end{figure}

Since stationarity does not hold for the extended time series with the same regularization window, we truncate the time series into segments, and investigate the stationarity of each period on a local scale. We use the COVID-19 pandemic as the flagging event for dividing the time series to explore the stationarity before and after COVID-19, given the higher market fluctuations observed post-pandemic. The initial impact of the pandemic on the US stock market was evident on Feb 24, 2020, when the S\&P 500 and Nasdaq declined by 3\% \cite{McCabe2020}. Therefore, we consider the time series from Jan 1996 to Feb 21st, 2020 (market closed on the weekend of 22-23 Feb) as the first period. The second period is from Feb 24th, 2020 to Dec 31st, 2023. For each segment, we apply the detrending time window $\Delta_1 t=1$ week, and the normalization window $\Delta_2 t$ is gradually narrowed down from 60 min to 40 min, 20 min, and 10 min. Results of the stationarity test for each time period are shown in Figure \ref{fig:SPX1996_2020} and Figure \ref{fig:SPX2020_2023}. In Figure \ref{fig:SPX1996_2020}, normalized price return before Covid presents stationarity with a regularization window at 60 mins, also at smaller window sizes, consistent with the results in the previous study. However, for the period after Covid, a gap is observed between $|\widehat{x^{**}} (f)|^2$ and $\widehat{C_{x^{**}}} (f)$ at $\Delta_2 t=60$ mins, indicating non-stationarity, \Blue{as shown in Figure \ref{fig:SPX2020_2023}-a}. This gap narrows as $\Delta_2 t$ decreases, and approaches a close-to-stationary result at $\Delta_2 t=10$ min, \Blue{as shown in Figure \ref{fig:SPX2020_2023}-d. This result of increasing level of stationarity with decreasing normalization window $\Delta_2 t$ in the post-Covid period is similar to what we observed in the full dataset of 28 years.}

\begin{figure}[!ht]
\centering
\includegraphics[scale=0.6, clip]{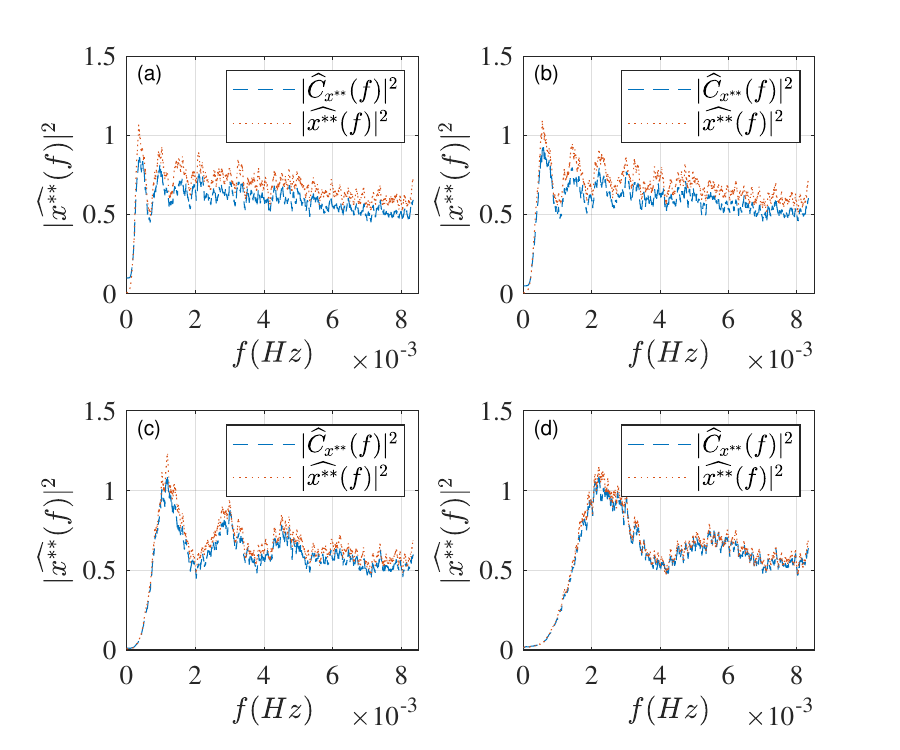}
\caption{Stationarity test with S\&P500 data from Feb 24st, 2020 to Dec 31st, 2020. (a) regularization window of 60 min. (b) regularization window of 40 min. (c) regularization window of 20 min. (d) regularization window of 10 min.}
\label{fig:SPX2020}     
\end{figure}

\begin{figure}[!ht]
\centering
\includegraphics[scale=0.6, clip]{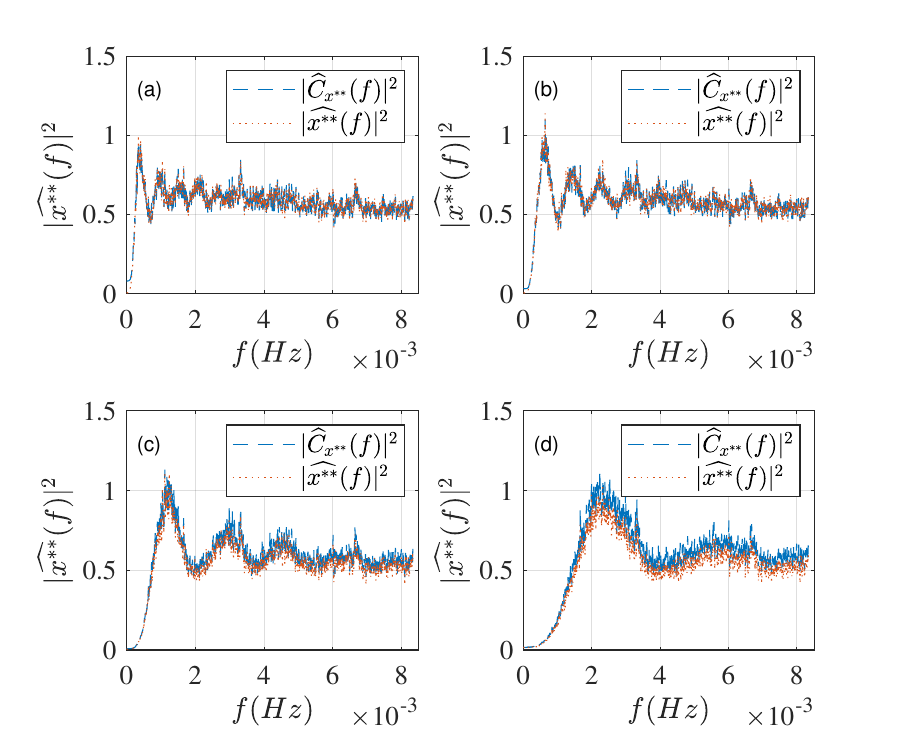}
\caption{Stationarity test with S\&P500 data from Jan 4th, 2021 to Dec 31st, 2023. (a) regularization window of 60 min. (b) regularization window of 40 min. (c) regularization window of 20 min. (d) regularization window of 10 min.}
\label{fig:SPX2021_2023}     
\end{figure}

As the normalized price return after Covid did not exhibit a satisfactory level of stationarity, we further divide this period into two segments. The time series from Feb 24th, 2020, to Dec 31st, 2020, is taken as a separate segment, representing the first year of coronavirus spread. The second segment spans from Jan 1st, 2021 to Dec 31st, 2023. Stationarity tests are performed on these two time periods following the same parameters of $\Delta_1 t$ and $\Delta_2 t$, and results are presented in Figure \ref{fig:SPX2020} and Figure \ref{fig:SPX2021_2023}, \Blue{respectively}. By further segmenting the time series into shorter periods, both periods achieve stationarity at a better level. \Blue{Compared to the data period from 2020 to 2023, where the normalized price return presents a close-to-stationary result with a small normalization window of $\Delta_2 t=10$ min, this shorter time period of the year 2020 present stationarity at $\Delta_2 t =20$ min, and an improved match between $|\widehat{x^{**}} (f)|^2$ and $\widehat{C_{x^{**}}} (f)$ is obtained at $\Delta_2 t=10$ min, as shown in Figure \ref{fig:SPX2020}-c and Figure \ref{fig:SPX2020}-d.} For the time period from 2021 to 2023, stationarity is achieved at $\Delta_2 t=60$ mins, without requiring a smaller size for $\Delta_2 t$. 

\begin{figure}[!ht]
\centering
\includegraphics[scale=0.6, clip]{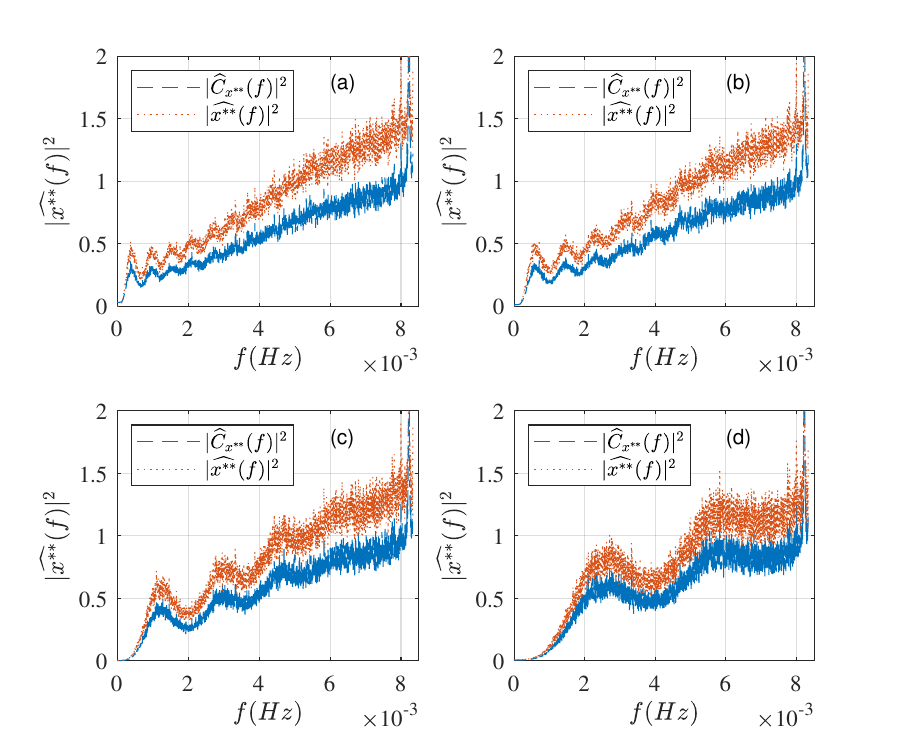}
\caption{Stationarity test with Bitcoin data from Apr 2nd, 2019 to Dec 31st, 2020. (a) regularization window of 60 min. (b) regularization window of 40 min. (c) regularization window of 20 min. (d) regularization window of 10 min.}
\label{fig:BTC_period1}     
\end{figure}

\begin{figure}[!ht]
\centering
\includegraphics[scale=0.6, clip]{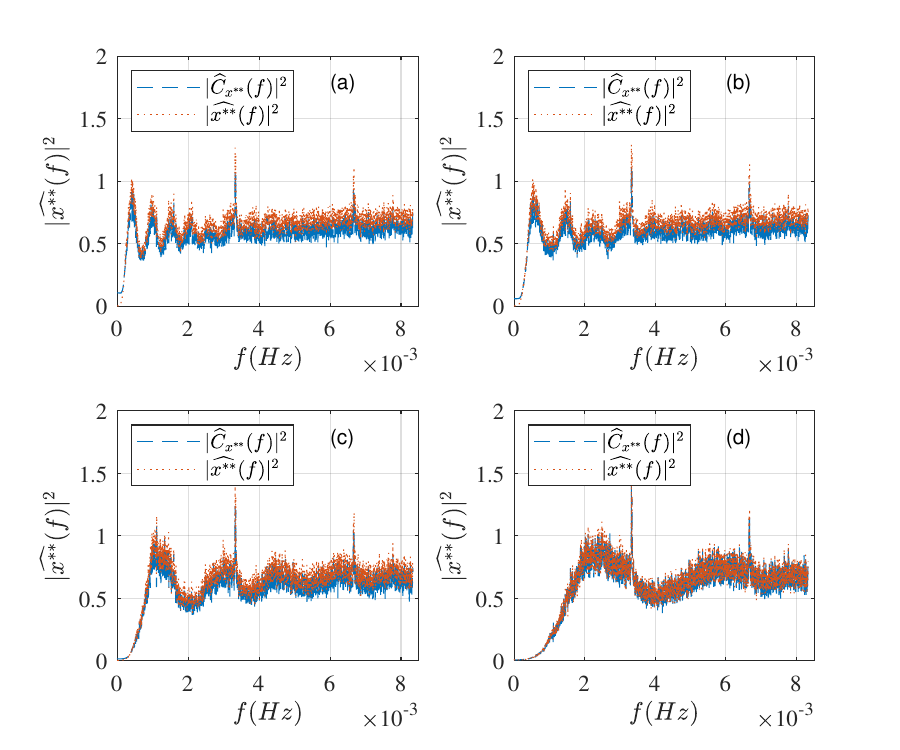}
\caption{Stationarity test with Bitcoin data from Jan 1st, 2021 to May 3rd, 2022. (a) regularization window of 60 min. (b) regularization window of 40 min. (c) regularization window of 20 min. (d) regularization window of 10 min.}
\label{fig:BTC_period2}     
\end{figure}

\begin{figure}[!ht]
\centering
\includegraphics[scale=0.6, clip]{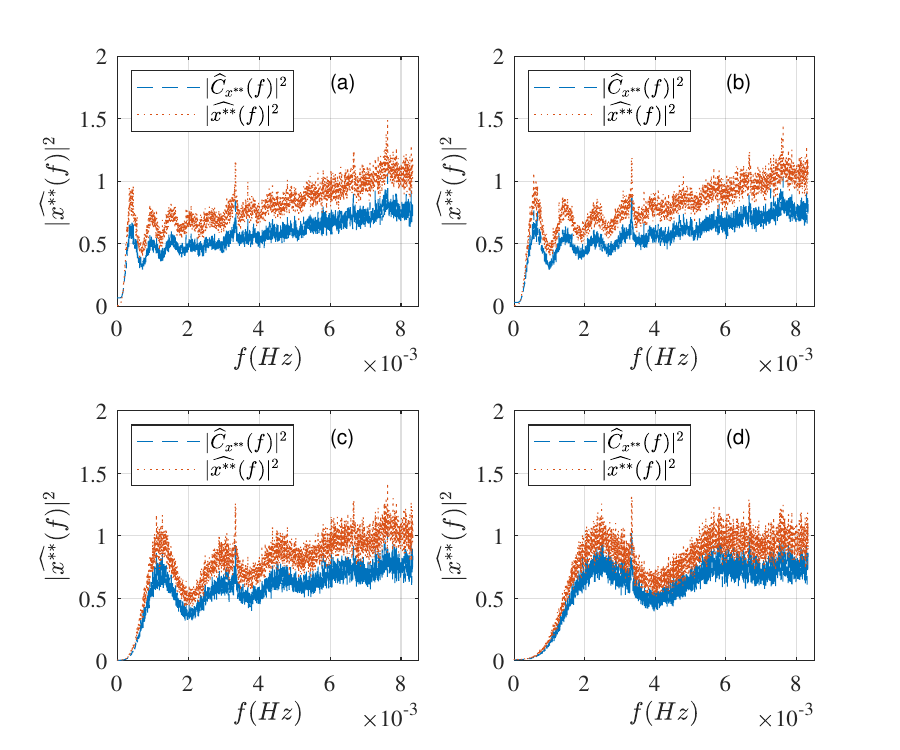}
\caption{Stationarity test with Bitcoin data from  May 4th, 2022 to Dec 31st, 2023. (a) regularization window of 60 min. (b) regularization window of 40 min. (c) regularization window of 20 min. (d) regularization window of 10 min.}
\label{fig:BTC_period3}     
\end{figure}

Testing results for the S\&P500 demonstrate global non-stationarity for the full 28-year dataset. However, from a local perspective, the normalized price returns are stationary. In the following part of this paper, we focus on testing the normalized price return of Bitcoin. In our previous examination of the statistical characteristics of Bitcoin price returns, we identified two time periods for Bitcoin \cite{tang2024stylized}. Here, we follow the same truncated time periods and extend the dataset to December 2023. The three segments for Bitcoin are: (1) Apr 2nd, 2019, to Dec 31st, 2020; (2) Jan 1st, 2021, to May 3rd, 2022; (3) May 4th, 2022, to Dec 31st, 2023. The price return is detrended with $\Delta_1 t=1$ week, and the normalization window of 60 min, 40 min, 20 min, and 10 min is applied to the respective time period. Results for each period are plotted in Figure \ref{fig:BTC_period1}, Figure \ref{fig:BTC_period2}, and Figure \ref{fig:BTC_period3}, respectively. For Bitcoin periods 1 and 3, \Blue{calculated $|\widehat{x^{**}} (f)|^2$ and $\widehat{C_{x^{**}}}$ for the detrended price return presents parallel lines at $\Delta_2 t=60$ min, indicating non-stationarity. Similar to the S\&P500,} the detrended price return requires a smaller regularization window of 10 min to achieve stationarity. In contrast, for period 2, during which the price experienced larger fluctuations, a stationary result can be obtained with $\Delta_2 t=60$ min.

\section{Discussion}

\textcolor{black}{In the earlier work of St\u{a}ric\u{a} and Granger~\cite{stuaricua2005nonstationarities} they carried out a detailed analysis of the period 1957 - 2000 as they had found this was a particularly `simple' interval in the history of the S\&P500. In this 47-year period they established that the absolute log-returns of the S\&P500 was an i.i.d., non-stationary random process with time-varying unconditional variance. The model is indeed quite simple, the absolute log-returns are given by a well-studied diffusion model:}
\begin{align} 
X(t) &= \mu(t) + \sigma(t) \epsilon_t
\end{align}
that had used a test statistic given by:
\begin{align}
    T &= \frac{X(t) - \mu(t)}{\sigma(t)}
\end{align}
\textcolor{black}{See eq. 17 in~\cite{stuaricua2005nonstationarities}, c.f. the standard score normalization of price returns in eq.~\ref{regularised} above.}

\Blue{Different from the work of St\u{a}ric\u{a} and Granger~\cite{stuaricua2005nonstationarities}, we standardized} \textcolor{black}{the distribution with an estimated time-varying mean $\mu(\Delta_1 t)$ and variance $\sigma(\Delta_2 t)$ where $\Delta_i t$ indicates the width of the time window over which these terms are estimated (recall that $\Delta_1 t$ removes the trend, while $\Delta_2 t$ attempts to make the time series WSP). This provides an extra degree of freedom over the St\u{a}ric\u{a} and Granger method while the data used provides a temporal resolution that was not possible using their inter-day data. What we have demonstrated here is a novel approach to understanding the `complexity' of a uni-variate time-series by seeking the maximum window size (if it exists) that allows the time-series to be brought into a stationary state: the shorter this window, the more heterogeneous the time series is. \Blue{Moreover, our results show that in some time periods, a relatively small window is required to achieve stationarity, which indicates the instability of the time series.} Further work will develop this approach into a more complete analytical methodology that we believe has shown here some considerable promise as a tool.}


\begin{thebibliography}{33}%
\makeatletter
\providecommand \@ifxundefined [1]{%
 \@ifx{#1\undefined}
}%
\providecommand \@ifnum [1]{%
 \ifnum #1\expandafter \@firstoftwo
 \else \expandafter \@secondoftwo
 \fi
}%
\providecommand \@ifx [1]{%
 \ifx #1\expandafter \@firstoftwo
 \else \expandafter \@secondoftwo
 \fi
}%
\providecommand \natexlab [1]{#1}%
\providecommand \enquote  [1]{``#1''}%
\providecommand \bibnamefont  [1]{#1}%
\providecommand \bibfnamefont [1]{#1}%
\providecommand \citenamefont [1]{#1}%
\providecommand \href@noop [0]{\@secondoftwo}%
\providecommand \href [0]{\begingroup \@sanitize@url \@href}%
\providecommand \@href[1]{\@@startlink{#1}\@@href}%
\providecommand \@@href[1]{\endgroup#1\@@endlink}%
\providecommand \@sanitize@url [0]{\catcode `\\12\catcode `\$12\catcode `\&12\catcode `\#12\catcode `\^12\catcode `\_12\catcode `\%12\relax}%
\providecommand \@@startlink[1]{}%
\providecommand \@@endlink[0]{}%
\providecommand \url  [0]{\begingroup\@sanitize@url \@url }%
\providecommand \@url [1]{\endgroup\@href {#1}{\urlprefix }}%
\providecommand \urlprefix  [0]{URL }%
\providecommand \Eprint [0]{\href }%
\providecommand \doibase [0]{http://dx.doi.org/}%
\providecommand \selectlanguage [0]{\@gobble}%
\providecommand \bibinfo  [0]{\@secondoftwo}%
\providecommand \bibfield  [0]{\@secondoftwo}%
\providecommand \translation [1]{[#1]}%
\providecommand \BibitemOpen [0]{}%
\providecommand \bibitemStop [0]{}%
\providecommand \bibitemNoStop [0]{.\EOS\space}%
\providecommand \EOS [0]{\spacefactor3000\relax}%
\providecommand \BibitemShut  [1]{\csname bibitem#1\endcsname}%
\let\auto@bib@innerbib\@empty
\bibitem [{\citenamefont {Cheng}\ \emph {et~al.}(2015)\citenamefont {Cheng}, \citenamefont {Sa-Ngasoongsong}, \citenamefont {Beyca}, \citenamefont {Le}, \citenamefont {Yang}, \citenamefont {Kong},\ and\ \citenamefont {Bukkapatnam}}]{cheng2015time}%
  \BibitemOpen
  \bibfield  {author} {\bibinfo {author} {\bibfnamefont {C.}~\bibnamefont {Cheng}}, \bibinfo {author} {\bibfnamefont {A.}~\bibnamefont {Sa-Ngasoongsong}}, \bibinfo {author} {\bibfnamefont {O.}~\bibnamefont {Beyca}}, \bibinfo {author} {\bibfnamefont {T.}~\bibnamefont {Le}}, \bibinfo {author} {\bibfnamefont {H.}~\bibnamefont {Yang}}, \bibinfo {author} {\bibfnamefont {Z.}~\bibnamefont {Kong}}, \ and\ \bibinfo {author} {\bibfnamefont {S.~T.}\ \bibnamefont {Bukkapatnam}},\ }\href@noop {} {\bibfield  {journal} {\bibinfo  {journal} {Iie Transactions}\ }\textbf {\bibinfo {volume} {47}},\ \bibinfo {pages} {1053} (\bibinfo {year} {2015})}\BibitemShut {NoStop}%
\bibitem [{\citenamefont {Aue}\ \emph {et~al.}(2015)\citenamefont {Aue}, \citenamefont {Norinho},\ and\ \citenamefont {H{\"o}rmann}}]{aue2015prediction}%
  \BibitemOpen
  \bibfield  {author} {\bibinfo {author} {\bibfnamefont {A.}~\bibnamefont {Aue}}, \bibinfo {author} {\bibfnamefont {D.~D.}\ \bibnamefont {Norinho}}, \ and\ \bibinfo {author} {\bibfnamefont {S.}~\bibnamefont {H{\"o}rmann}},\ }\href@noop {} {\bibfield  {journal} {\bibinfo  {journal} {Journal of the American Statistical Association}\ }\textbf {\bibinfo {volume} {110}},\ \bibinfo {pages} {378} (\bibinfo {year} {2015})}\BibitemShut {NoStop}%
\bibitem [{\citenamefont {Jentsch}\ and\ \citenamefont {Rao}(2015)}]{jentsch2015test}%
  \BibitemOpen
  \bibfield  {author} {\bibinfo {author} {\bibfnamefont {C.}~\bibnamefont {Jentsch}}\ and\ \bibinfo {author} {\bibfnamefont {S.~S.}\ \bibnamefont {Rao}},\ }\href@noop {} {\bibfield  {journal} {\bibinfo  {journal} {Journal of Econometrics}\ }\textbf {\bibinfo {volume} {185}},\ \bibinfo {pages} {124} (\bibinfo {year} {2015})}\BibitemShut {NoStop}%
\bibitem [{\citenamefont {Dahlhaus}(2012)}]{dahlhaus2012locally}%
  \BibitemOpen
  \bibfield  {author} {\bibinfo {author} {\bibfnamefont {R.}~\bibnamefont {Dahlhaus}},\ }in\ \href@noop {} {\emph {\bibinfo {booktitle} {Handbook of statistics}}},\ Vol.~\bibinfo {volume} {30}\ (\bibinfo  {publisher} {Elsevier},\ \bibinfo {year} {2012})\ pp.\ \bibinfo {pages} {351--413}\BibitemShut {NoStop}%
\bibitem [{\citenamefont {Bhowmik}\ and\ \citenamefont {Wang}(2020)}]{bhowmik2020stock}%
  \BibitemOpen
  \bibfield  {author} {\bibinfo {author} {\bibfnamefont {R.}~\bibnamefont {Bhowmik}}\ and\ \bibinfo {author} {\bibfnamefont {S.}~\bibnamefont {Wang}},\ }\href@noop {} {\bibfield  {journal} {\bibinfo  {journal} {Entropy}\ }\textbf {\bibinfo {volume} {22}},\ \bibinfo {pages} {522} (\bibinfo {year} {2020})}\BibitemShut {NoStop}%
\bibitem [{\citenamefont {Harvey}\ \emph {et~al.}(2016)\citenamefont {Harvey}, \citenamefont {Leybourne}, \citenamefont {Sollis},\ and\ \citenamefont {Taylor}}]{harvey2016tests}%
  \BibitemOpen
  \bibfield  {author} {\bibinfo {author} {\bibfnamefont {D.~I.}\ \bibnamefont {Harvey}}, \bibinfo {author} {\bibfnamefont {S.~J.}\ \bibnamefont {Leybourne}}, \bibinfo {author} {\bibfnamefont {R.}~\bibnamefont {Sollis}}, \ and\ \bibinfo {author} {\bibfnamefont {A.~R.}\ \bibnamefont {Taylor}},\ }\href@noop {} {\bibfield  {journal} {\bibinfo  {journal} {Journal of Empirical Finance}\ }\textbf {\bibinfo {volume} {38}},\ \bibinfo {pages} {548} (\bibinfo {year} {2016})}\BibitemShut {NoStop}%
\bibitem [{\citenamefont {Jin}(2017)}]{jin2017time}%
  \BibitemOpen
  \bibfield  {author} {\bibinfo {author} {\bibfnamefont {X.}~\bibnamefont {Jin}},\ }\href@noop {} {\bibfield  {journal} {\bibinfo  {journal} {International Review of Economics \& Finance}\ }\textbf {\bibinfo {volume} {51}},\ \bibinfo {pages} {157} (\bibinfo {year} {2017})}\BibitemShut {NoStop}%
\bibitem [{\citenamefont {Onnela}\ \emph {et~al.}(2003{\natexlab{a}})\citenamefont {Onnela}, \citenamefont {Chakraborti}, \citenamefont {Kaski}, \citenamefont {Kertesz},\ and\ \citenamefont {Kanto}}]{onnela2003asset}%
  \BibitemOpen
  \bibfield  {author} {\bibinfo {author} {\bibfnamefont {J.-P.}\ \bibnamefont {Onnela}}, \bibinfo {author} {\bibfnamefont {A.}~\bibnamefont {Chakraborti}}, \bibinfo {author} {\bibfnamefont {K.}~\bibnamefont {Kaski}}, \bibinfo {author} {\bibfnamefont {J.}~\bibnamefont {Kertesz}}, \ and\ \bibinfo {author} {\bibfnamefont {A.}~\bibnamefont {Kanto}},\ }\href@noop {} {\bibfield  {journal} {\bibinfo  {journal} {Physica Scripta}\ }\textbf {\bibinfo {volume} {2003}},\ \bibinfo {pages} {48} (\bibinfo {year} {2003}{\natexlab{a}})}\BibitemShut {NoStop}%
\bibitem [{\citenamefont {Onnela}\ \emph {et~al.}(2003{\natexlab{b}})\citenamefont {Onnela}, \citenamefont {Chakraborti}, \citenamefont {Kaski},\ and\ \citenamefont {Kertesz}}]{onnela2003dynamic}%
  \BibitemOpen
  \bibfield  {author} {\bibinfo {author} {\bibfnamefont {J.-P.}\ \bibnamefont {Onnela}}, \bibinfo {author} {\bibfnamefont {A.}~\bibnamefont {Chakraborti}}, \bibinfo {author} {\bibfnamefont {K.}~\bibnamefont {Kaski}}, \ and\ \bibinfo {author} {\bibfnamefont {J.}~\bibnamefont {Kertesz}},\ }\href@noop {} {\bibfield  {journal} {\bibinfo  {journal} {Physica A: Statistical Mechanics and its Applications}\ }\textbf {\bibinfo {volume} {324}},\ \bibinfo {pages} {247} (\bibinfo {year} {2003}{\natexlab{b}})}\BibitemShut {NoStop}%
\bibitem [{\citenamefont {Onnela}\ \emph {et~al.}(2003{\natexlab{c}})\citenamefont {Onnela}, \citenamefont {Chakraborti}, \citenamefont {Kaski}, \citenamefont {Kertesz},\ and\ \citenamefont {Kanto}}]{onnela2003dynamics}%
  \BibitemOpen
  \bibfield  {author} {\bibinfo {author} {\bibfnamefont {J.-P.}\ \bibnamefont {Onnela}}, \bibinfo {author} {\bibfnamefont {A.}~\bibnamefont {Chakraborti}}, \bibinfo {author} {\bibfnamefont {K.}~\bibnamefont {Kaski}}, \bibinfo {author} {\bibfnamefont {J.}~\bibnamefont {Kertesz}}, \ and\ \bibinfo {author} {\bibfnamefont {A.}~\bibnamefont {Kanto}},\ }\href@noop {} {\bibfield  {journal} {\bibinfo  {journal} {Physical Review E}\ }\textbf {\bibinfo {volume} {68}},\ \bibinfo {pages} {056110} (\bibinfo {year} {2003}{\natexlab{c}})}\BibitemShut {NoStop}%
\bibitem [{\citenamefont {Harr{\'e}}\ and\ \citenamefont {Bossomaier}(2009)}]{harre2009phase}%
  \BibitemOpen
  \bibfield  {author} {\bibinfo {author} {\bibfnamefont {M.}~\bibnamefont {Harr{\'e}}}\ and\ \bibinfo {author} {\bibfnamefont {T.}~\bibnamefont {Bossomaier}},\ }\href@noop {} {\bibfield  {journal} {\bibinfo  {journal} {Europhysics Letters}\ }\textbf {\bibinfo {volume} {87}},\ \bibinfo {pages} {18009} (\bibinfo {year} {2009})}\BibitemShut {NoStop}%
\bibitem [{\citenamefont {Harr{\'e}}(2015)}]{harre2015entropy}%
  \BibitemOpen
  \bibfield  {author} {\bibinfo {author} {\bibfnamefont {M.}~\bibnamefont {Harr{\'e}}},\ }in\ \href@noop {} {\emph {\bibinfo {booktitle} {Proceedings of the International Conference on Social Modeling and Simulation, plus Econophysics Colloquium 2014}}}\ (\bibinfo {organization} {Springer International Publishing},\ \bibinfo {year} {2015})\ pp.\ \bibinfo {pages} {15--25}\BibitemShut {NoStop}%
\bibitem [{\citenamefont {Harr{\'e}}(2022)}]{harre2022entropy}%
  \BibitemOpen
  \bibfield  {author} {\bibinfo {author} {\bibfnamefont {M.~S.}\ \bibnamefont {Harr{\'e}}},\ }\href@noop {} {\bibfield  {journal} {\bibinfo  {journal} {Entropy}\ }\textbf {\bibinfo {volume} {24}},\ \bibinfo {pages} {210} (\bibinfo {year} {2022})}\BibitemShut {NoStop}%
\bibitem [{\citenamefont {Harr{\'e}}\ and\ \citenamefont {Zaitouny}(2023)}]{harre2023detecting}%
  \BibitemOpen
  \bibfield  {author} {\bibinfo {author} {\bibfnamefont {M.~S.}\ \bibnamefont {Harr{\'e}}}\ and\ \bibinfo {author} {\bibfnamefont {A.}~\bibnamefont {Zaitouny}},\ }\href@noop {} {\bibfield  {journal} {\bibinfo  {journal} {Expert Systems with Applications}\ }\textbf {\bibinfo {volume} {211}},\ \bibinfo {pages} {118437} (\bibinfo {year} {2023})}\BibitemShut {NoStop}%
\bibitem [{\citenamefont {Gardner}(1978)}]{gardner1978stationarizable}%
  \BibitemOpen
  \bibfield  {author} {\bibinfo {author} {\bibfnamefont {W.}~\bibnamefont {Gardner}},\ }\href@noop {} {\bibfield  {journal} {\bibinfo  {journal} {IEEE Transactions on information theory}\ }\textbf {\bibinfo {volume} {24}},\ \bibinfo {pages} {8} (\bibinfo {year} {1978})}\BibitemShut {NoStop}%
\bibitem [{\citenamefont {Chen}\ \emph {et~al.}(2002)\citenamefont {Chen}, \citenamefont {Ivanov}, \citenamefont {Hu},\ and\ \citenamefont {Stanley}}]{Chen2002effect}%
  \BibitemOpen
  \bibfield  {author} {\bibinfo {author} {\bibfnamefont {Z.}~\bibnamefont {Chen}}, \bibinfo {author} {\bibfnamefont {P.~C.}\ \bibnamefont {Ivanov}}, \bibinfo {author} {\bibfnamefont {K.}~\bibnamefont {Hu}}, \ and\ \bibinfo {author} {\bibfnamefont {H.~E.}\ \bibnamefont {Stanley}},\ }\href {\doibase 10.1103/PhysRevE.65.041107} {\bibfield  {journal} {\bibinfo  {journal} {Phys. Rev. E}\ }\textbf {\bibinfo {volume} {65}},\ \bibinfo {pages} {041107} (\bibinfo {year} {2002})}\BibitemShut {NoStop}%
\bibitem [{\citenamefont {Tsallis}(1988)}]{Tsallis1988Possible}%
  \BibitemOpen
  \bibfield  {author} {\bibinfo {author} {\bibfnamefont {C.}~\bibnamefont {Tsallis}},\ }\href {\doibase 10.1007/BF01016429} {\bibfield  {journal} {\bibinfo  {journal} {Journal of Statistical Physics}\ }\textbf {\bibinfo {volume} {52}},\ \bibinfo {pages} {479} (\bibinfo {year} {1988})}\BibitemShut {NoStop}%
\bibitem [{\citenamefont {Borland}(2002)}]{borland2002option}%
  \BibitemOpen
  \bibfield  {author} {\bibinfo {author} {\bibfnamefont {L.}~\bibnamefont {Borland}},\ }\href@noop {} {\bibfield  {journal} {\bibinfo  {journal} {Physical review letters}\ }\textbf {\bibinfo {volume} {89}},\ \bibinfo {pages} {098701} (\bibinfo {year} {2002})}\BibitemShut {NoStop}%
\bibitem [{\citenamefont {Witt}\ \emph {et~al.}(1998)\citenamefont {Witt}, \citenamefont {Kurths},\ and\ \citenamefont {Pikovsky}}]{witt1998testing}%
  \BibitemOpen
  \bibfield  {author} {\bibinfo {author} {\bibfnamefont {A.}~\bibnamefont {Witt}}, \bibinfo {author} {\bibfnamefont {J.}~\bibnamefont {Kurths}}, \ and\ \bibinfo {author} {\bibfnamefont {A.}~\bibnamefont {Pikovsky}},\ }\href@noop {} {\bibfield  {journal} {\bibinfo  {journal} {physical Review E}\ }\textbf {\bibinfo {volume} {58}},\ \bibinfo {pages} {1800} (\bibinfo {year} {1998})}\BibitemShut {NoStop}%
\bibitem [{\citenamefont {Last}\ and\ \citenamefont {Shumway}(2008)}]{last2008detecting}%
  \BibitemOpen
  \bibfield  {author} {\bibinfo {author} {\bibfnamefont {M.}~\bibnamefont {Last}}\ and\ \bibinfo {author} {\bibfnamefont {R.}~\bibnamefont {Shumway}},\ }\href@noop {} {\bibfield  {journal} {\bibinfo  {journal} {Journal of multivariate analysis}\ }\textbf {\bibinfo {volume} {99}},\ \bibinfo {pages} {191} (\bibinfo {year} {2008})}\BibitemShut {NoStop}%
\bibitem [{\citenamefont {Vincent}\ \emph {et~al.}(2010)\citenamefont {Vincent}, \citenamefont {Giebel}, \citenamefont {Pinson},\ and\ \citenamefont {Madsen}}]{vincent2010resolving}%
  \BibitemOpen
  \bibfield  {author} {\bibinfo {author} {\bibfnamefont {C.}~\bibnamefont {Vincent}}, \bibinfo {author} {\bibfnamefont {G.}~\bibnamefont {Giebel}}, \bibinfo {author} {\bibfnamefont {P.}~\bibnamefont {Pinson}}, \ and\ \bibinfo {author} {\bibfnamefont {H.}~\bibnamefont {Madsen}},\ }\href@noop {} {\bibfield  {journal} {\bibinfo  {journal} {Journal of Applied Meteorology and Climatology}\ }\textbf {\bibinfo {volume} {49}},\ \bibinfo {pages} {253} (\bibinfo {year} {2010})}\BibitemShut {NoStop}%
\bibitem [{\citenamefont {Hegger}\ \emph {et~al.}(2000)\citenamefont {Hegger}, \citenamefont {Kantz}, \citenamefont {Matassini},\ and\ \citenamefont {Schreiber}}]{hegger2000coping}%
  \BibitemOpen
  \bibfield  {author} {\bibinfo {author} {\bibfnamefont {R.}~\bibnamefont {Hegger}}, \bibinfo {author} {\bibfnamefont {H.}~\bibnamefont {Kantz}}, \bibinfo {author} {\bibfnamefont {L.}~\bibnamefont {Matassini}}, \ and\ \bibinfo {author} {\bibfnamefont {T.}~\bibnamefont {Schreiber}},\ }\href@noop {} {\bibfield  {journal} {\bibinfo  {journal} {Physical Review Letters}\ }\textbf {\bibinfo {volume} {84}},\ \bibinfo {pages} {4092} (\bibinfo {year} {2000})}\BibitemShut {NoStop}%
\bibitem [{\citenamefont {Lim}\ \emph {et~al.}(2023)\citenamefont {Lim}, \citenamefont {Park}, \citenamefont {Kim}, \citenamefont {Wi}, \citenamefont {Lim}, \citenamefont {Jeon}, \citenamefont {Choi},\ and\ \citenamefont {Park}}]{lim2023long}%
  \BibitemOpen
  \bibfield  {author} {\bibinfo {author} {\bibfnamefont {S.}~\bibnamefont {Lim}}, \bibinfo {author} {\bibfnamefont {J.}~\bibnamefont {Park}}, \bibinfo {author} {\bibfnamefont {S.}~\bibnamefont {Kim}}, \bibinfo {author} {\bibfnamefont {H.}~\bibnamefont {Wi}}, \bibinfo {author} {\bibfnamefont {H.}~\bibnamefont {Lim}}, \bibinfo {author} {\bibfnamefont {J.}~\bibnamefont {Jeon}}, \bibinfo {author} {\bibfnamefont {J.}~\bibnamefont {Choi}}, \ and\ \bibinfo {author} {\bibfnamefont {N.}~\bibnamefont {Park}},\ }in\ \href@noop {} {\emph {\bibinfo {booktitle} {2023 IEEE International Conference on Big Data (BigData)}}}\ (\bibinfo {organization} {IEEE},\ \bibinfo {year} {2023})\ pp.\ \bibinfo {pages} {748--757}\BibitemShut {NoStop}%
\bibitem [{\citenamefont {Arias-Calluari}\ \emph {et~al.}(2022)\citenamefont {Arias-Calluari}, \citenamefont {Najafi}, \citenamefont {Harr{\'e}}, \citenamefont {Tang},\ and\ \citenamefont {Alonso-Marroquin}}]{arias2022testing}%
  \BibitemOpen
  \bibfield  {author} {\bibinfo {author} {\bibfnamefont {K.}~\bibnamefont {Arias-Calluari}}, \bibinfo {author} {\bibfnamefont {M.~N.}\ \bibnamefont {Najafi}}, \bibinfo {author} {\bibfnamefont {M.~S.}\ \bibnamefont {Harr{\'e}}}, \bibinfo {author} {\bibfnamefont {Y.}~\bibnamefont {Tang}}, \ and\ \bibinfo {author} {\bibfnamefont {F.}~\bibnamefont {Alonso-Marroquin}},\ }\href {\doibase https://doi.org/10.1016/j.physa.2021.126487} {\bibfield  {journal} {\bibinfo  {journal} {Physica A: Statistical Mechanics and its Applications}\ }\textbf {\bibinfo {volume} {587}},\ \bibinfo {pages} {126487} (\bibinfo {year} {2022})}\BibitemShut {NoStop}%
\bibitem [{\citenamefont {Arias-Calluari}\ \emph {et~al.}(2018)\citenamefont {Arias-Calluari}, \citenamefont {Alonso-Marroquin},\ and\ \citenamefont {Harr{\'e}}}]{arias2018closed}%
  \BibitemOpen
  \bibfield  {author} {\bibinfo {author} {\bibfnamefont {K.}~\bibnamefont {Arias-Calluari}}, \bibinfo {author} {\bibfnamefont {F.}~\bibnamefont {Alonso-Marroquin}}, \ and\ \bibinfo {author} {\bibfnamefont {M.~S.}\ \bibnamefont {Harr{\'e}}},\ }\href@noop {} {\bibfield  {journal} {\bibinfo  {journal} {Physical Review E}\ }\textbf {\bibinfo {volume} {98}},\ \bibinfo {pages} {012103} (\bibinfo {year} {2018})}\BibitemShut {NoStop}%
\bibitem [{\citenamefont {Alonso-Marroquin}\ \emph {et~al.}(2019)\citenamefont {Alonso-Marroquin}, \citenamefont {Arias-Calluari}, \citenamefont {Harr{\'e}}, \citenamefont {Najafi},\ and\ \citenamefont {Herrmann}}]{alonso2019q}%
  \BibitemOpen
  \bibfield  {author} {\bibinfo {author} {\bibfnamefont {F.}~\bibnamefont {Alonso-Marroquin}}, \bibinfo {author} {\bibfnamefont {K.}~\bibnamefont {Arias-Calluari}}, \bibinfo {author} {\bibfnamefont {M.}~\bibnamefont {Harr{\'e}}}, \bibinfo {author} {\bibfnamefont {M.~N.}\ \bibnamefont {Najafi}}, \ and\ \bibinfo {author} {\bibfnamefont {H.~J.}\ \bibnamefont {Herrmann}},\ }\href@noop {} {\bibfield  {journal} {\bibinfo  {journal} {Physical Review E}\ }\textbf {\bibinfo {volume} {99}},\ \bibinfo {pages} {062313} (\bibinfo {year} {2019})}\BibitemShut {NoStop}%
\bibitem [{\citenamefont {Arias-Calluari}\ \emph {et~al.}(2021)\citenamefont {Arias-Calluari}, \citenamefont {Alonso-Marroquin}, \citenamefont {Najafi},\ and\ \citenamefont {Harr{\'e}}}]{arias2021methods}%
  \BibitemOpen
  \bibfield  {author} {\bibinfo {author} {\bibfnamefont {K.}~\bibnamefont {Arias-Calluari}}, \bibinfo {author} {\bibfnamefont {F.}~\bibnamefont {Alonso-Marroquin}}, \bibinfo {author} {\bibfnamefont {M.~N.}\ \bibnamefont {Najafi}}, \ and\ \bibinfo {author} {\bibfnamefont {M.}~\bibnamefont {Harr{\'e}}},\ }\href@noop {} {\bibfield  {journal} {\bibinfo  {journal} {Physica A: Statistical Mechanics and its Applications}\ }\textbf {\bibinfo {volume} {568}},\ \bibinfo {pages} {125587} (\bibinfo {year} {2021})}\BibitemShut {NoStop}%
\bibitem [{\citenamefont {Tang}\ \emph {et~al.}(2024)\citenamefont {Tang}, \citenamefont {Arias-Calluari}, \citenamefont {Harré},\ and\ \citenamefont {Alonso-Marroquin}}]{tang2024stylized}%
  \BibitemOpen
  \bibfield  {author} {\bibinfo {author} {\bibfnamefont {Y.}~\bibnamefont {Tang}}, \bibinfo {author} {\bibfnamefont {K.}~\bibnamefont {Arias-Calluari}}, \bibinfo {author} {\bibfnamefont {M.~S.}\ \bibnamefont {Harré}}, \ and\ \bibinfo {author} {\bibfnamefont {F.}~\bibnamefont {Alonso-Marroquin}},\ }\href@noop {} {\enquote {\bibinfo {title} {Stylized facts of high-frequency bitcoin time series},}\ } (\bibinfo {year} {2024}),\ \Eprint {http://arxiv.org/abs/2402.11930} {arXiv:2402.11930 [q-fin.ST]} \BibitemShut {NoStop}%
\bibitem [{\citenamefont {St{\u{a}}ric{\u{a}}}\ and\ \citenamefont {Granger}(2005)}]{stuaricua2005nonstationarities}%
  \BibitemOpen
  \bibfield  {author} {\bibinfo {author} {\bibfnamefont {C.}~\bibnamefont {St{\u{a}}ric{\u{a}}}}\ and\ \bibinfo {author} {\bibfnamefont {C.}~\bibnamefont {Granger}},\ }\href@noop {} {\bibfield  {journal} {\bibinfo  {journal} {Review of economics and statistics}\ }\textbf {\bibinfo {volume} {87}},\ \bibinfo {pages} {503} (\bibinfo {year} {2005})}\BibitemShut {NoStop}%
\bibitem [{\citenamefont {Box}\ \emph {et~al.}(2015)\citenamefont {Box}, \citenamefont {Jenkins}, \citenamefont {Reinsel},\ and\ \citenamefont {Ljung}}]{box2015time}%
  \BibitemOpen
  \bibfield  {author} {\bibinfo {author} {\bibfnamefont {G.~E.}\ \bibnamefont {Box}}, \bibinfo {author} {\bibfnamefont {G.~M.}\ \bibnamefont {Jenkins}}, \bibinfo {author} {\bibfnamefont {G.~C.}\ \bibnamefont {Reinsel}}, \ and\ \bibinfo {author} {\bibfnamefont {G.~M.}\ \bibnamefont {Ljung}},\ }\href@noop {} {\emph {\bibinfo {title} {Time series analysis: forecasting and control}}}\ (\bibinfo  {publisher} {John Wiley \& Sons},\ \bibinfo {year} {2015})\BibitemShut {NoStop}%
\bibitem [{\citenamefont {Liu}\ \emph {et~al.}(1999)\citenamefont {Liu}, \citenamefont {Gopikrishnan}, \citenamefont {Stanley} \emph {et~al.}}]{liu1999statistical}%
  \BibitemOpen
  \bibfield  {author} {\bibinfo {author} {\bibfnamefont {Y.}~\bibnamefont {Liu}}, \bibinfo {author} {\bibfnamefont {P.}~\bibnamefont {Gopikrishnan}}, \bibinfo {author} {\bibfnamefont {H.~E.}\ \bibnamefont {Stanley}},  \emph {et~al.},\ }\href@noop {} {\bibfield  {journal} {\bibinfo  {journal} {Physical review e}\ }\textbf {\bibinfo {volume} {60}},\ \bibinfo {pages} {1390} (\bibinfo {year} {1999})}\BibitemShut {NoStop}%
\bibitem [{\citenamefont {Schafer}(2011)}]{schafer2011savitzky}%
  \BibitemOpen
  \bibfield  {author} {\bibinfo {author} {\bibfnamefont {R.~W.}\ \bibnamefont {Schafer}},\ }\href@noop {} {\bibfield  {journal} {\bibinfo  {journal} {IEEE Signal processing magazine}\ }\textbf {\bibinfo {volume} {28}},\ \bibinfo {pages} {111} (\bibinfo {year} {2011})}\BibitemShut {NoStop}%
\bibitem [{\citenamefont {McCabe}(2020)}]{McCabe2020}%
  \BibitemOpen
  \bibfield  {author} {\bibinfo {author} {\bibfnamefont {C.}~\bibnamefont {McCabe}},\ }\href {https://www.wsj.com/articles/stocks-fall-as-coronavirus-spread-accelerates-outside-china-11582533308} {\enquote {\bibinfo {title} {Dow industrials close 1,000 points lower as coronavirus cases mount},}\ } (\bibinfo {year} {2020})\BibitemShut {NoStop}%
\end{thebibliography}
\end{document}